\documentclass[aps,preprint,showpacs,preprintnumbers,amsmath,amssymb,superscriptaddress]{revtex4}

\usepackage{graphicx}

\begin{document}

\title{Dynamical hysteresis in multisite protein modification.}
\author{Edoardo~Milotti}
\affiliation{Dipartimento di Fisica, Universit\`{a} di Trieste,
and INFN -- Sezione di Trieste, Via Valerio, 2, I-34127 Trieste, Italy}
\email{milotti@ts.infn.it}
\author{Alessio Del Fabbro}
\affiliation{Istituto Nazionale di Fisica Nucleare, Sezione di Trieste, Via Valerio 2, I-34127 Trieste, Italy}
\author{Chiara~Dalla~Pellegrina}
\affiliation{Dipartimento Scientifico e Tecnologico, Universit\`a di Verona, Strada Le Grazie 15 Ð CV1, I-37134 Verona, Italy}
\author{Roberto~Chignola}
\affiliation{Dipartimento Scientifico e Tecnologico, Universit\`a di Verona, Strada Le Grazie 15 Ð CV1, I-37134 Verona, Italy}
\affiliation{Istituto Nazionale di Fisica Nucleare, Sezione di Trieste, Via Valerio 2, I-34127 Trieste, Italy}

\date{\today}

\begin{abstract}
Multisite protein modification is a ubiquitous mechanism utilized by cells to control protein functions. We have recently proposed a dynamical description of multisite protein modification which embodies all the essential features of the process (E. Milotti, A. Del Fabbro, C. Dalla Pellegrina, and R. Chignola, Physica A, in press), and we have used this model to analyze the stability and the time-scales of this mechanism. The same model can be used to understand how the system responds to stimuli: here we show that it displays frequency-dependent dynamical hysteresis. This behavior closely parallels -- with the due differences -- what is observed in magnetic systems. By selecting model parameters that span the known biological ranges, we find that the  frequency-dependent features cover the band of the observed oscillations of molecular intracellular signals, and this suggests that this mechanism may have an important role in cellular information processing.
\end{abstract}

\maketitle


Multisite protein phosphorylation, nitrosylation, methylation, etc., are very common mechanisms in cells, which use these processes to control protein functions \cite{yang}, and thus the flow of information in biochemical networks. We have recently introduced a dynamical model of multisite protein modification which describes many properties of these ubiquitous mechanisms \cite{ourpap2}. The basic idea is that a molecule $B$ can dock to anyone of several sites on protein $A$: this produces structural changes in the protein, and eventually, when enough sites are occupied by $B$, the protein undergoes a conformational switch which modifies its chemical activity or causes the release of some other substance. In this way it is possible to model threshold processes in cells, and the thresholds turn out to be stable and tunable over many orders of magnitude. This mechanism is actually quite complex and in \cite{ourpap2} we use simplifying assumptions that lead to the following differential system for the concentrations:
\begin{eqnarray}
\nonumber
\frac{d[A_{0}]}{dt} & = & -N k_{+} [A_{0}] [B] + k_{-} [A_{1}]\\
\nonumber
&&\ldots\\
\nonumber
\frac{d[A_{n}]}{dt} & = & -n k_{-} [A_{n}] -(N-n) k_{+} [A_{n}] [B] \\
\nonumber
&&+(N-n+1) k_{+} [A_{n-1}] [B] + (n+1) k_{-} [A_{n+1}]\\
\label{gensys}
&&\ldots\\
\nonumber
\frac{d[A_{N}]}{dt} & = & -N k_{-} [A_{N}] + k_{+} [A_{N-1}] [B] 
\end{eqnarray}
where the states $A_n$ are the forms of $A$ with exactly $n$ modified sites out of a total of $N$ sites, and $k_+$ and $k_-$ are the on-off rates. This is a non-dissipative differential system, and in  \cite{ourpap2} we have studied its dynamical properties, we have extended it to include the modified form of $A$ produced by threshold crossing, and finally found that the bimolecular attachment/detachment process produces a dynamics that is the same as the classical allosteric effect.
In this paper we return to the unmodified system (\ref{gensys}), and turn our attention to other features of the process. 

\begin{figure}
\includegraphics[width=3.2in]{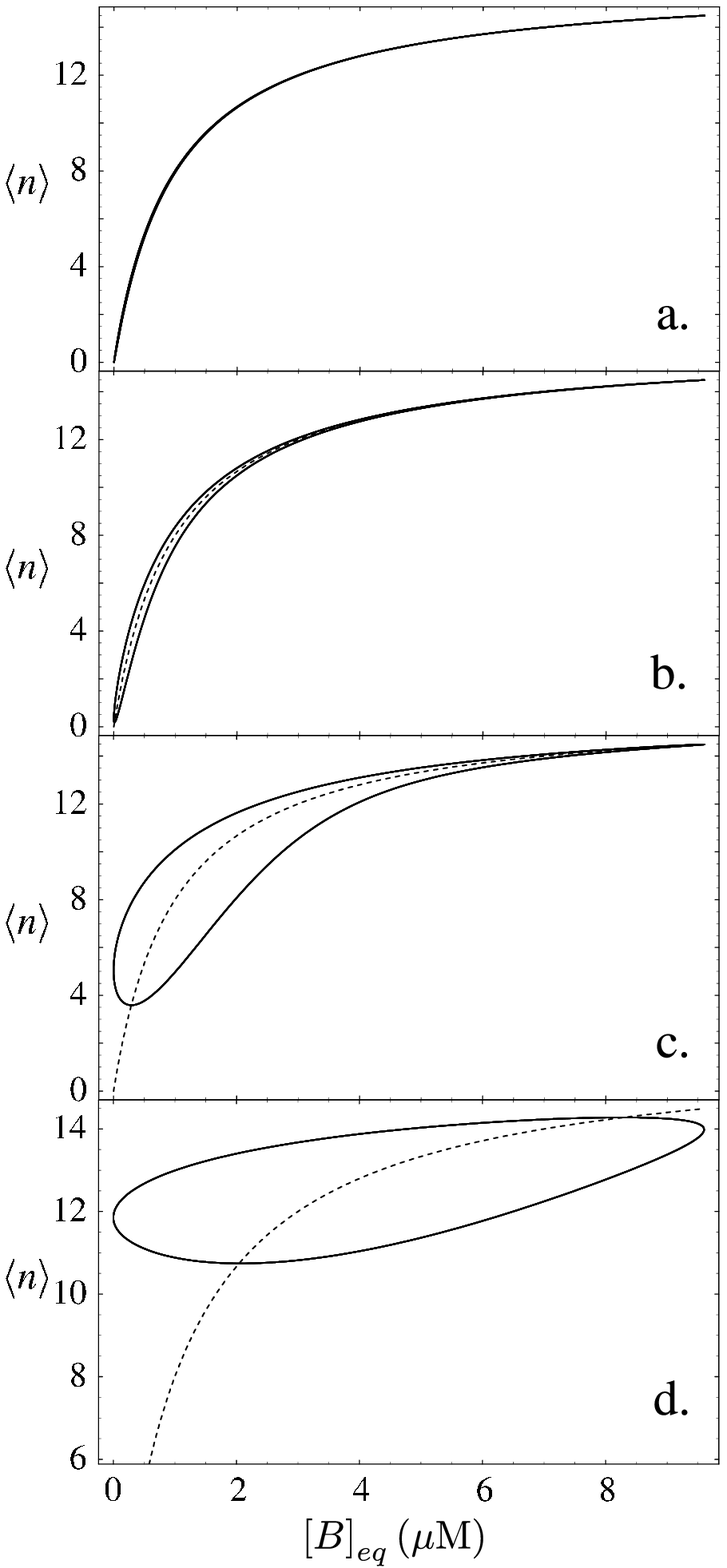}
\caption{\label{fig1} Plot of the average number of modified sites $\langle n \rangle$ vs. $[B]_{eq}$ with system parameters $N=16$, $[A]_0 = 10 \mu$M,  $k_- =$ 1 Hz, $k_+ = 10^6$ Hz M$^{-1}$. The dotted curve is the static solution (\ref{neq}). The loops correspond to different modulation frequencies $\nu$: {\bf a.} $\nu=1$ mHz;  {\bf b.} $\nu=10$ mHz; {\bf c.} $\nu = 100$ mHz; {\bf d.} $\nu= 1$ Hz. The system follows each loop counterclockwise.}
\end{figure}

\begin{figure}
\includegraphics[width=3.2in]{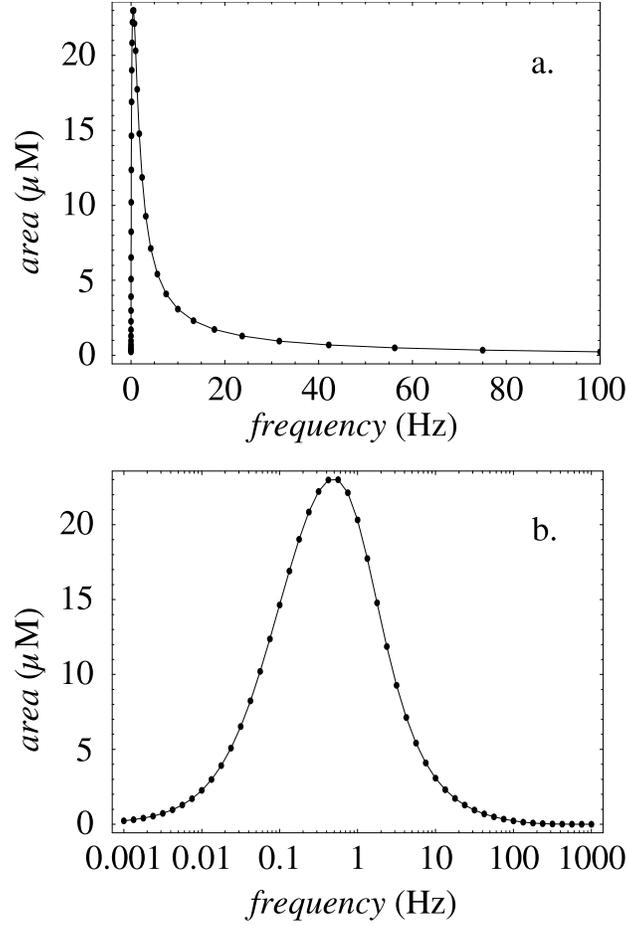}
\caption{\label{fig2}  This figure shows the loop area vs. modulation frequency $\nu$  for $N=16$, $[A]_0 = 10 \mu$M,  $k_- =$ 1 Hz, $k_+ = 10^6$ Hz M$^{-1}$. The dots represent the individual numerical integrations of the differential system. The plot with logarithmic horizontal scale in part {\bf b.} shows that the curve is close to a log-normal shape \cite{ws}.}
\end{figure}

\begin{figure}
\includegraphics[width=3.2in]{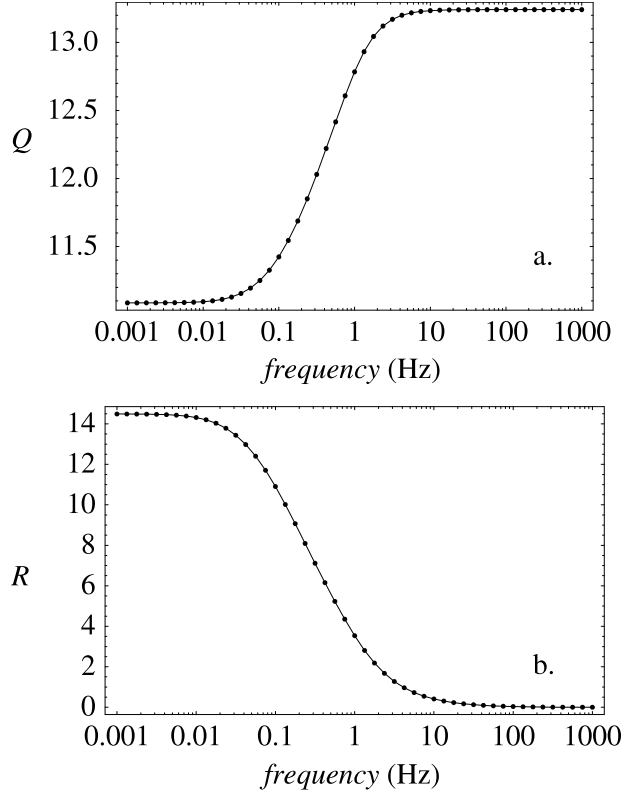}
\caption{\label{fig3}  Plot {\bf a.} of the order parameter $Q$, and {\bf b.} of the vertical range $R$ vs. modulation frequency $\nu$ for $N=16$, $[A]_0 = 10 \mu$M,  $k_- =$ 1 Hz, $k_+ = 10^6$ Hz M$^{-1}$.}
\end{figure}

\begin{figure}
\includegraphics[width=3.2in]{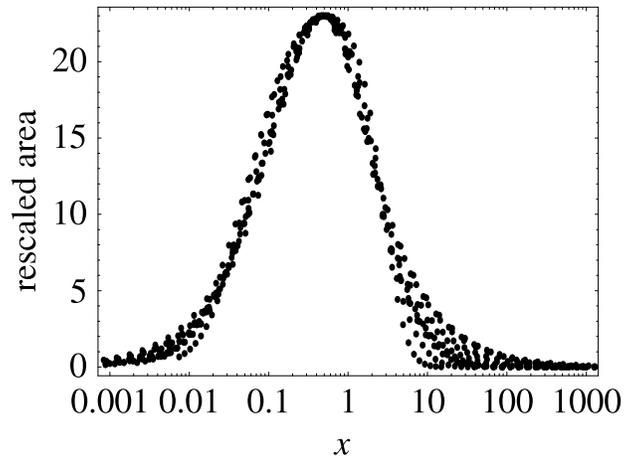}
\caption{\label{fig4}  Rescaled loop area (arb. units) vs. rescaled frequency $x$ for several different parameter sets. The rescaled frequency is $x = (\nu/\nu_0)^\alpha$, where the index $\alpha$  is proportional to $1/\sigma_0^2$.}
\end{figure}

In \cite{ourpap2} we assumed that the concentration of $B$ behaves quasi-statically, so that the following conservation equation also holds: 
\begin{equation}
\label{Bcons}
\sum_{n=1}^{N} n[A_n] + [B] = [B]_0
\end{equation}
where $[B]_0$ is the total concentration of $B$. Equation  (\ref{Bcons}) means that $B$ can either be found free in the solution or bound to $A$, and that each $A_n$ counts $n$ times as much, because $n$ molecules $B$ are bound to it. However equation (\ref{Bcons}) is not adequate to describe a situation in which $B$ is driven by an external process. At first sight it may seem that the differential equation 
\begin{equation}
\label{wrong}
\frac{d[B]}{dt} = -\sum_{n=1}^{N} n\frac{[A_n]}{dt}  +\frac{d[B]_0}{dt}
\end{equation}
where $d[B]_0/dt$ is a driving term, solves the problem. However this is not so, because this equation can easily push $[B]$ towards unphysical negative values. The problem is that $B$ is partly bound to $A$, and one might drive the total amount of $B$ down to small values so fast that the bound $B$ does not have the time to unbind. 
Biology provides an easy way out: the concentration of many proteins is commonly modulated by the interplay of production and destruction; production is triggered by genetic expression, while destruction may be activated by the expression of a molecule that finally leads to the ubiquitination of the target protein, and its degradation by proteasomes.
It is easy to see that if the ubiquitination process followed by proteasome action is fast enough, the details of the destruction process can be neglected, and the change of $[B]$ due to simultaneous production (P) and destruction (D) can be described by the equation 
\begin{equation}
\label{pdeq}
\left.\frac{d[B]}{dt}\right|_{P,D} = v_B(t) - k_U [B]
\end{equation}
where $v_B(t)$ is the production rate and $k_U$ is the decay (destruction) constant associated to ubiquitination. When we revise (\ref{wrong}) to include (\ref{pdeq}), we find 
\begin{equation}
\label{Bequation}
\frac{d[B]}{dt} = -\sum_{n=1}^{N} n\frac{[A_n]}{dt}  +  v_B - k_U [B]
\end{equation}
Here we use the production rate $v_B(t)$ as the driving term that modulates the concentration of $B$: we let $v_B(t)$ change periodically
\begin{equation}
\label{vrate}
v_B(t) = v_{B,0}\left[1+m \sin\left(2 \pi \nu t\right) \right]
\end{equation}
where $m$ is the modulation index and $\nu$ is the modulation frequency of the production process. This modulation only roughly approximates the actual biological ups and downs in enzyme production, but has the obvious advantage that the differential system is excited by a single Fourier component, and thus the nonlinearities that are observed in the response are unambiguous properties of the differential system.
The differential system cannot be solved analytically, and we have to resort to numerical integration methods, and explore the parameter space as well as we can. On the whole there are eight parameters, i.e., the total number of modification sites $N$, the total amount of protein $A$, the on-off rates $k_+$ and $k_-$, the production rate $v_{B,0}$ along with the modulation index $m$  and frequency $\nu$, and the decay rate $k_U$. At first sight the exploration of such a large parameter space might seem to be a daunting task, but fortunately we can limit ourselves to the restricted ranges of the actual biological systems, and we use as starting points some of the values associated with the multisite phosphorylation of the retinoblastoma protein (Rb) (as in \cite{ourpap2,ourpap1}), i.e., $N=16$, $[A]_0 = \sum_{n=1}^{N} [A_n] = 10 \mu$M,  $k_- =$ 1 Hz, $k_+ = 10^6$ Hz M$^{-1}$. Using reasonable values for ubiquitin concentration and for the forward rate in the Michaelis-Menten reaction for ubiquitination, we find that $k_U$ is in the range 1-10000 Hz. From equation (\ref{Bequation}) we see that under stationary conditions, $v_B=k_U [B]$, and since we know \cite{ourpap1} that  $B$ has a critical breakpoint at $[B] = N[A]_0$, we take $v_B$ in the range $(0,2 N k_U [A]_0)$. The modulation index $m$ can only range from 0 to 1, and here we take the full amplitude $m=1$. Finally, from our previous knowledge of the eigenvalue distribution for the linearized system studied in \cite{ourpap2}, we know that the relaxation times in the multisite phosphorylation of the Rb protein are approximately in the range 1 ms -- 100 s, and therefore -- after we have fixed all the other parameters -- we sweep this range with uniform logarithmic spacing.
In the simplified situation that we have chosen to highlight the dynamical properties of multisite modification, it is natural to monitor the average number of modified sites  $\langle n \rangle = \sum_{n=1}^{N} n[A_n]/[A]_0$; using equation (\ref{vrate}) we see that the the static concentration of $B$ should be 
\begin{equation}
\label{Brate}
[B]_{eq} = \frac{v_{B,0}}{k_U}\left[1+m \sin\left(2 \pi \nu t\right) \right]
\end{equation}
and correspondingly, we expect to find the average number of modified sites \cite{ourpap2}:
\begin{equation}
\label{neq}
\langle n\rangle_{eq}=\frac{N (k_+/k_-) [B]_{eq}}{1+(k_+/k_-) [B]_{eq}}
\end{equation}
Figure \ref{fig1} shows that this expectation is fulfilled only at very low frequencies: for frequencies higher than a few mHz, the curve (\ref{neq}) opens up into a loop that is reminiscent of the hysteresis loop in magnetic systems. Here the loop cannot be symmetric about the origin, because both concentration and the average number of modified sites are bound to be non-negative, however, just like in magnetic systems \cite{chach} the loop shape changes and at the same time the centroid of the loop migrates to a different position at high frequency. We can define a dynamic order parameter $Q$ as in \cite{chach}: 
\begin{equation}
Q = \frac{1}{T} \oint \langle n\rangle  dt
\end{equation}
which is just $\langle n\rangle$ averaged over the hysteresis loop, however here it is also useful to consider the vertical range $R =  \langle n\rangle_{max} - \langle n\rangle_{min}$. Figures \ref{fig2} and figure \ref{fig3} show the loop area, the order parameter $Q$, and the vertical range $R$ vs. frequency, for the standard choice of parameters listed above. Figure \ref{fig2} shows that the area behaves roughly as a log-normal function, and therefore the high-frequency tail has an approximate power-law dependence \cite{ws}. This kind of behavior does not change appreciably when the other adjustable parameters are varied as explained above. 
The loop area in magnetic systems in known to follow a scaling law \cite{chach}: here we find that the loop area obeys the approximate scaling law
\begin{equation}
A(\nu) \approx C \exp\left\{-\frac{\left[\ln(\nu/\nu_0)\right]^{2}}{2\sigma^2_0} \right\}
\end{equation}
where the coefficients $C$, $\nu_0$, $\sigma_0$ depend on the specific simulation parameters, and therefore the tail of the distribution has an approximate $1/\nu$ dependence. The coefficients also seem to follow simple scaling laws that depend on $v_{B,0}$: 
\begin{eqnarray}
C(v_{B,0}) & = & C_0 \exp\left\{ -\frac{[\ln((v_{B,0})/v_C)]^2}{2 \sigma_C^2}  \right\}\\
\sigma_0(v_{B,0}) & = & a + b v_{B,0}^w \\
\nu_0 & = & d + e v_{B,0}^r
\end{eqnarray}
The new coefficients $C_0$, $v_C$, $\sigma_C$, $a$, $b$, $w$, $d$, $e$ and $r$ depend on the remaining system parameters $N$, $[A]_0$, $k_+$, $k_-$, and $k_U$. Figure \ref{fig4} shows the superposition of several plots of rescaled loop area obtained in different numerical integration runs: the superposition is reasonably good, although more work shall be needed to obtain a better scaling law and to fix the association between  $C$, $\nu_0$, $\sigma_0$ and the simulation parameters. 

\begin{figure}
\includegraphics[width=3.2in]{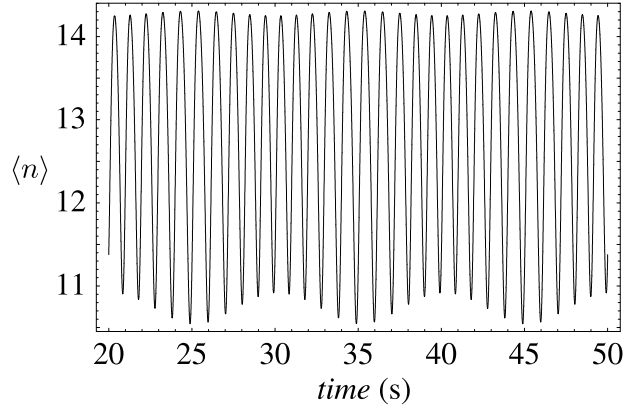}
\caption{\label{fig5}  Example of amplitude-modulated output obtained from a frequency-modulated fixed-amplitude input. In this case the system parameters are as before $N=16$, $[A]_0 = 10 \mu$M,  $k_- =$ 1 Hz, $k_+ = 10^6$ Hz M$^{-1}$, and $\nu_C$ = 1 Hz, $\nu_M$ = 0.1 Hz, $m$ = 0.8. }
\end{figure}

The vertical range $R$ in figure \ref{fig3} shows that the system behaves as a low pass filter with an extended transition region, and since this behavior emerges in a biological setting it is natural to wonder whether it may have a deeper meaning. We conjecture that it could be used in some biochemical circuits as a kind of slope filter, i.e., a low-pass filter used in frequency modulation decoding that converts frequency modulation to amplitude modulation (see, e.g., \cite{fm}). Figure \ref{fig5} shows the result of a calculation with a frequency-modulated $B$ production rate
\begin{equation}
\label{vrateFM}
v_B(t) = v_{B,0}\left\{1+\sin\left[2 \pi \nu_C t + m \sin \left( 2 \pi \nu_M t \right) \right] \right\}
\end{equation}
where $\nu_C$ is a fixed frequency, $\nu_M$ is the modulation frequency, and $m$ is again a modulation index. Although the system parameters have not been optimized for this purpose, we see that indeed a frequency-modulated, fixed-amplitude input produces an amplitude-modulated output. 

The importance of hysteresis in enzyme kinetics was first stressed by C. Frieden almost 40 years ago \cite{cf}:  here we have shown how dynamical hysteresis and related effects arise naturally in a ubiquitous biological mechanism, multisite protein modification. Hysteresis shows up in many biomolecular pathways, and it has been observed experimentally in reactions that involve proteins with multiple phosphorylation sites, e.g., the phosphatase Cdc25 \cite{bhyst}. Phosphatase Cdc25 regulates cell division, and it is believed that hysteresis underlies the irreversibility of the cell-cycle transition into and out of mitosis. 
We also note that information stored in environmental molecules can propagate in the cell through the oscillations of various types of intracellular molecules (such as c-AMP, ATP, NADPH, Ca$^{2+}$) and that various important cell functions (e.g., insulin secretion, immune cell activation, neuron transmission, gene expression, etc.) are activated by the frequency decoding of this information. The Ca$^{2+}$ oscillations are a paradigmatic instance, and several intracellular proteins/enzymes can bind Ca$^{2+}$ at multiple sites and act as frequency decoders \cite{cara}. One of these is CaM Kinase II whose enzymatic activity has been experimentally shown to be sensitive to the frequency of Ca$^{2+}$ oscillations \cite{deko}. Most importantly, the observed periods of intracellular Ca$^{2+}$ oscillations range between a few seconds to minutes, and match quite well the frequency range that we obtain with realistic system parameters (the transition in figure \ref{fig3} spans the range 0.01 Hz - 10 Hz).

\pagebreak



\end{document}